\pdfoutput=1

\documentclass[letterpaper, 10 pt, conference]{ieeeconf}  

\IEEEoverridecommandlockouts

\usepackage{cite}
\usepackage{lipsum} 
\usepackage[T1]{fontenc}
\usepackage{amsmath}
\usepackage{newtxtext}
\usepackage[cmintegrals]{newtxmath}
\usepackage{bm} 
\usepackage{topcapt}
\usepackage{footnote}
\usepackage{threeparttablex}
\usepackage[pdftex]{graphicx}  
\usepackage{booktabs}
\usepackage{soul}

\IEEEoverridecommandlockouts        

\usepackage{tikz}
\usepackage{textcomp}
\usepackage{lipsum}

\title{\LARGE \bf
  Can We Monitor Breathing During Sleep via Wi-Fi on Smartphone?
}

\author{Alexander Tataraidze, \IEEEmembership{Member, IEEE},  Roman Olesyuk, and Mikhail Pikhletsky
\thanks{A. Tataraidze is with Huawei Technologies and Bauman Moscow State Technical University, Moscow, Russia (tataraidze.alexander@huawei.com).}%
\thanks{R. Olesyuk and M. Pikhletsky are with Huawei Technologies, 127562, Moscow, Russia, Altufevskoe h., 1 (roman.olesyuk@huawei.com, pikhletsky.mikhail@huawei.com).}%
}

\begin{document}

\onecolumn
This is a preprint copy that has been accepted for publication in Proceedings of the 41th Annual International Conference of the IEEE Engineering in Medicine and Biology Society.

\bigskip 

\copyright  2019  IEEE.  Personal  use  of  this  material  is  permitted.  Permission  from  IEEE  must obtained  for  all  other  uses,  in  any current or future media, including reprinting/republishing this material for advertising or promotional purposes, creating new collective  works, for resale or redistribution to servers or lists, or reuse of any copyrighted component of this work in other works.

\newpage\twocolumn


\maketitle
\thispagestyle{empty}
\pagestyle{empty}

\begin{abstract}
   Over the last decade there has been high interest in home sleep monitoring among research community and society.  Smartphone-based sleep monitoring seems to be especially attractive. However, smartphone should be able to register physiological processes for that, such as breathing or heartbeats. In this study, we tried to detect breathing based on the analysis of channel state information (CSI) of Wi-Fi connection between a smartphone and an access point. We collected data of 5 subjects with different body mass index (17--33) in base sleep postures: prone, supine, and side. The obtained CSI-based respiratory signal was compared with ground truth from a contact belt sensor.  The mean correlation coefficient between the signals of 0.85 was achieved. The mean absolute difference between respiratory rates (breaths per epoch) is 0.39. The results show that it is possible to use smartphones without additional sensors for registering breathing based on CSI and potentially for sleep monitoring.  
\end{abstract}

\section{Introduction}

Over the last decade there has been an increasing interest in home sleep monitoring among the general population: estimation of sleep stages became a mandatory feature for fitness trackers, app stores have tens of application for sleep estimation~\cite{Ong2016a}. Apparently, the reasons are high prevalence of chronic sleep deprivation and sleep disorders. Around \mbox{30--40\%} of adults report sleep dissatisfaction~\cite{Kerkhof2017, Simonelli2018}. Obstructive sleep apnea affects more than 9\% of general population~\cite{Senaratna2017}, while the prevalence of insomnia is around 6--9\%~\cite{Kerkhof2017,Ohayon2002a}.

The gold standard for sleep estimation is polysomnography (PSG), when electroencephalography, electromyography and electrooculography are used for sleep staging. However, a number of studies have shown that it is possible to detect sleep stages based on the analysis of Heart Rate Variability (HRV) or respiratory patterns~\cite{Redmond2006, Long2014, Tataraidze2016, Tataraidze2017, Beattie2017}. Furthermore, it is possible to distinguish sleep and wakefulness episodes via analysis of movements. The method is called actigraphy, and it is widely used in sleep medicine.  

If we take a look to consumer devices market, we can see that most fitness trackers have a sleep monitoring function. They use photoplethysmography and accelerometer sensors to get HRV and estimate physical activity. There are also non-contact solutions, e.g. ResMed S+ monitors respiratory movements based on radio waves reflected from body, while Emfit and Beddit use ballistocardiography sensors to detect heartbeats and breathing. However, the necessity to buy a specialized device holds back the spread of home sleep monitoring technologies. It would be quite interesting to use already built-in smartphone equipment for this purpose. Recently, it has been shown that it is possible to monitor human breathing based on the analysis of \mbox{Wi-Fi} connection~\cite{Liu2015, Li2017, Gu2018a}.

Wi-Fi is a packet-based wireless transmission system and uses on orthogonal frequency-division multiplexing method. Currently widely adopted version is IEEE 802.11ac. It uses high bandwidth radio channels (from 20 MHz to 160 MHz) on 2.4 GHz and 5 GHz bands. The wide channel experiences frequency selective attenuations and thus a channel equalization is necessary. Here the channel consists of everything between digital-to-analog converter at the transmitter and analog-to-digital converter (ADC) at the receiver including radio frequency (RF) front ends, antennas at the transmitter and the receiver and all the media in between. We can treat the channel as a linear system and equalize it by dividing input signal after ADC in the frequency domain by the frequency response of the channel. 

In Wi-Fi systems the channel frequency response is called channel state information (CSI). CSI is estimated on the Wi-Fi chip every time a data packet is received and used internally for equalization.  CSI reflects a state of the media as well as the RF front ends and antennas. The RF front ends and antennas are designed to be stable over time so the most of the changes in CSI are due to changes in the media: changes of the reflection patterns and the attenuation properties caused by the movements. Thus, respiratory motions should have a predominant effect on CSI during sleep, if there are no other sources of movement between the connected devices.

\begin{figure*}
	\centering
	\includegraphics[width=\textwidth]{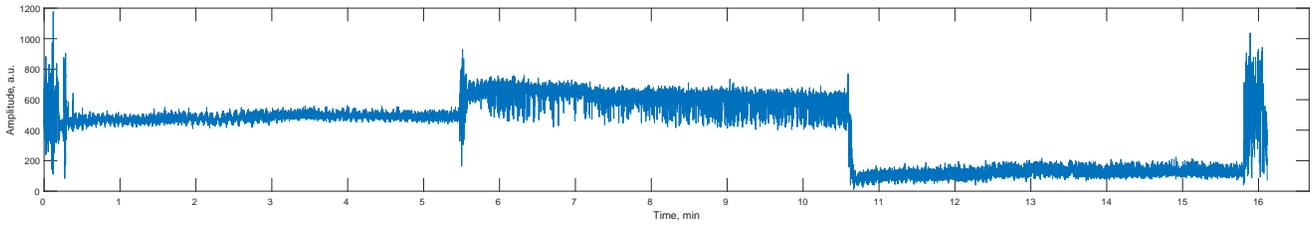}
	\caption{The row CSI signal for one subcarrier during the whole record. The subject was in supine position from 0.5 to 5.5 min, lying on his right side from 5.5 to 10.5 min, and in prone posture from 11 to 16 min. }
	\label{csi_long}
\end{figure*}

Antennas are a part of the channel and influence CSI significantly. A radiation pattern and a polarization are the most important properties of the antenna as well as a size. The simplest antenna design is a half-wavelength dipole. This antennas have a high gain and smooth omnidirectional radiation pattern and are very often used in the stationary equipment (e.g. Wi-Fi access points) where the size is not that important~\cite{roman6}. For the half-wavelength dipole the gain and the phase do not depend on the angle of arrival (AoA) in the horizontal plane. But this design is not suitable for the mobile devices due to the size limitation. 

More compact planar inverted F-antennas or planar monopole antennas are used in the smartphones~\cite{roman7}. While these antennas are small they have a low gain and an uneven radiation pattern which in turn decreases a sensitivity and creates directional changes in CSI because the gain and the phase now depend on the AoA~\cite{roman10, roman12}.

To the best of our knowledge, smartphones have not been used previously in studies devoted to breathing monitoring based on the analysis of CSI. However, it might be more challenging task to register breathing on smartphones than on other devices due to design and size of antennas. The aim of this paper is to check whether it is possible to monitor breathing during sleep via Wi-Fi CSI in conditions close to the real use case --- using a smartphone on a bedside table and a few meters away access point (AP). The study is consists of 2 parts. In the first one (\ref{signal_comparion}), we checked if CSI have information about respiratory motions generally in this scenario. In the second (\ref{cycle_detection}), we estimated quality of automated detection of respiratory cycles based on CSI in comparison to a belt sensor.

\section{Materials and Methods}

\subsection{Data collection}
 A smartphone was placed on a bedside table. We used Nexus 5 with Broadcom BCM4339 Wi-Fi IEEE 802.11ac chip and one antenna. The special firmware~\cite{Gu2018a} was applied to get access to CSI. As far as we know, currently it is the only one open firmware, which allows to have access to CSI values on mobile phones. The smartphone worked in passive mode only for receiving Wi-Fi packets. 
 
 As a transmitting AP we used a mini PC with Atheros AR9380 Wi-Fi NIC and one external antenna. The AP was located 6 m away from the opposite side of the bed. Wi-Fi signal transmitted at 5500 MHz with 20 MHz bandwidth. The delay between sending Wi-Fi packets was 16 ms. This value was selected due to limitations of the firmware for CSI extraction on the phone. So, the sampling frequency was around 62.5 Hz with some jitter.
  
 As ground truth we used a Vernier respiration monitor belt. It consists of a gas pressure sensor and a cuff with air fixed on the chest by a strap. The sensor monitors gas pressure changes caused by respiratory motions. The sampling frequency was 100~Hz.
 
 We collected data of 5 male subjects (mean age of 34, range of 28--36, mean Body Mass Index (BMI) of 26, range of 17--33). The subjects imitated sleep --- lay on the bed without moving and breathing in a comfortable rhythm --- during 5 min in each of three postures: supine, on the right side (the back was directed to the phone), and prone.

 \subsection{Signal Comparison}\label{signal_comparion}
 
  \begin{figure}[h!]
 	\centering
 	\includegraphics[width=\columnwidth]{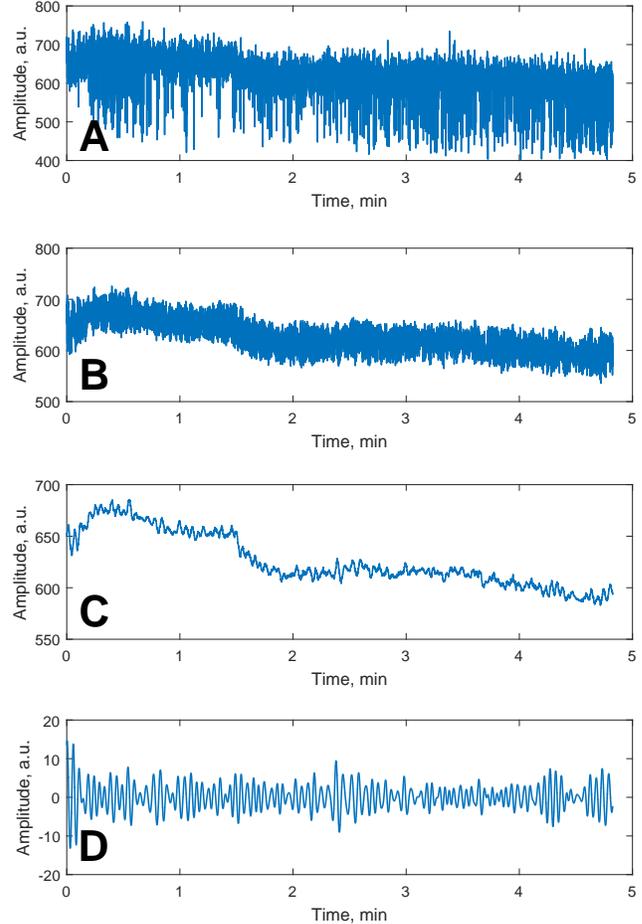}
 	\caption{The CSI subcarrier during processing steps: A --- the raw signal (CSI magnitude); B --- the signal after Hampel filter; C --- the signal after moving average filter; D --- the signal after Butterworth filter.}
 	\label{csi_short}
 \end{figure}

\begin{figure*}[t!]
	\centering
	\includegraphics[width=\textwidth]{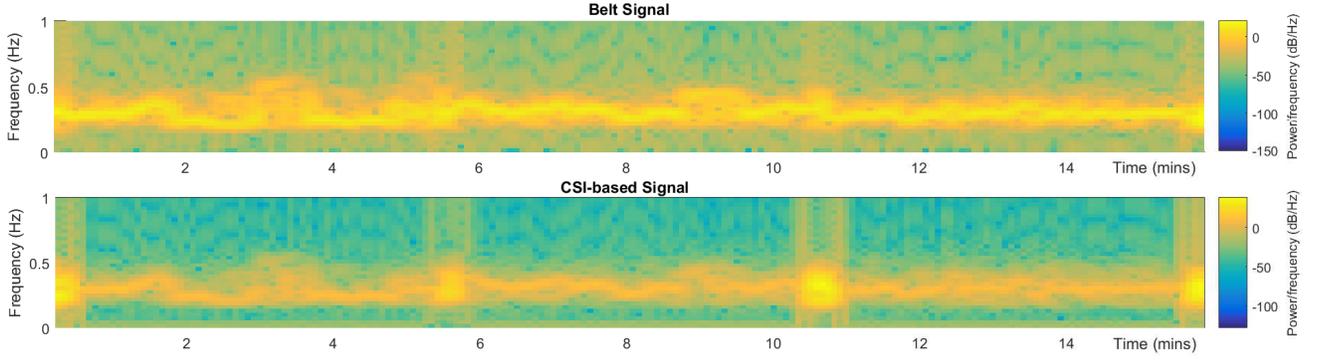}
	\caption{Spectrogram comparison between the belt signal and the CSI-based respiratory signal for one of the records}
	\label{spectrograms}
\end{figure*}

\begin{figure}[t!]
	\centering
	\includegraphics[width=\columnwidth]{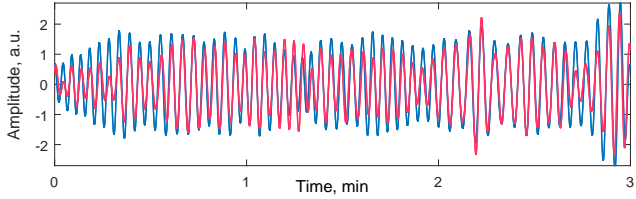}
	\caption{Comparison the CSI-based respiratory signal  with the corresponded belt signal}
	\label{csi_final}
\end{figure}

 The firmware provides CSI values for 50 subcarriers. CSI values for one subcarrier during the whole records are shown on Fig.~\ref{csi_long}. The preprocessing of CSI signals consists of the  following steps (\mbox{Fig.~\ref{csi_short}}):
 \begin{enumerate}
 	\item obtaining magnitude of CSI values;
 	\item removing outliers by the Hampel identifier~\cite{Davies2011,Liu2015} with a window of 1 s and a threshold of 1.7;
	\item linear interpolation to fit 60 Hz sampling frequency;
 	\item low-pass filtering by a moving average filter with a window of 1.5 s;
 	\item band-pass filtering by the 4th order Butterworth filter with cut-off frequencies of 0.2 Hz and 0.4 Hz.
 \end{enumerate}

The belt signal was resampled from 100 Hz to 60 Hz and band-pass filtered in the same range. The CSI record and the belt signal were manually synchronized based on reference points produced by specific phenomena (e.g. posture changing). 

Since the CSI record consists of 50 signal recorded simultaneously, we need to select the signal, which better represents information about breathing. The optimal subcarrier might be altered due to channel state changes. We used a window of 10 s and correlation with the belt signal as a criteria for channel selection. The signals were Z-normalized before correlation coefficient estimation. 

Mean correlation coefficient were calculated for each subject and the entire dataset. Moreover, visual spectrogram comparison between the signals were performed. 

\subsection{Automated Detection of Respiratory Cycles Based on CSI}\label{cycle_detection}

For the second step of the study we follow the same methodology for CSI signal preprocessing.  Channel selection was made with window of 30 s and Principal Component Analysis (PCA)~\cite{Pearson1901, Liu2015}. The length of the window corresponds to the standard epoch length in sleep analysis. The first component of PCA was used as the respiratory signal. Respiratory cycles were detected based on the analysis of turning points on the obtained signal and the correspond interval of the belt signal. Respiratory rates (RR) were calculated per epoch for both signals. Mean Absolute Difference (MAD) between these RR was estimated for each posture and subject, as well as percent of epochs, which have difference in RR. MAD was computed as

\begin{equation}
MAD = \dfrac{\sum\limits_{i=1}^{n} |N_{i}^{belt} - N_{i}^{CSI}|}{n},
\end{equation}
in which $n$ --- total number of epochs, $N_{i}^{belt}$ --- number of respiratory cycles in $i$-th epoch based on the analysis of the belt signal, $N_{i}^{CSI}$ --- number of respiratory cycles in $i$-th epoch based on the analysis of the CSI record.
 
\section{Results}

According to the methodology described in paragraph~\ref{signal_comparion}, the mean correlation coefficient of 0.85 (range of 0.78--0.92) were achieved between the belt signal and the CSI-based respiratory signal. 

Visual comparison of spectrograms (Fig.~\ref{spectrograms})  shows high correlation between respiration related spectral components (yellow "strips" in spectrograms) of belt and CSI signals (e.g. segments 1--2 min, 7.5--8.5 min, 11.5--12.5 min). Fig.~\ref{csi_final} shows the signals in the time domain.

\begin{figure}[b!]
	\centering
	\includegraphics[width=\columnwidth]{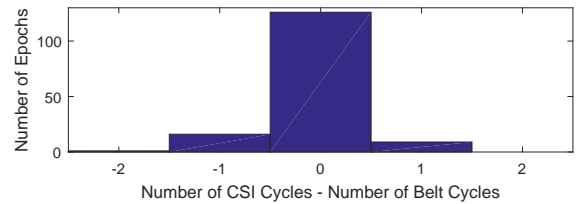}
	\caption{Difference between number of detected respiratory cycles on the CSI-based respiratory signal and the belt signal}
	\label{diff_cycles}
\end{figure}

MAD of 0.39 between RR were achieved (Table~\ref{table_error}). A third of the epochs have at least one wrong detected or missed cycle (Table~\ref{table_bad_epochs_1}), and 5\% of epochs have two or more (Table~\ref{table_bad_epochs_2}). Fig.~\ref{diff_cycles} shows a distribution of errors in RR estimation.

\begin{table}[t]
	\caption{Percent of Epochs with One or More Wrong Detected/Missed Respiratory Cycles}
	\label{table_bad_epochs_1}
	\begin{center}
		\begin{threeparttable}[c]
			\begin{tabular}{c c c c c}
				\hline
				Subject & Supine, \% & Side, \% & Prone, \% & All Postures, \%\\
				\hline
				1  & 40 & 60 & 70 & 57 \\			
				2  & 20 & 0  & 10 & 10 \\
				3  & 50 & 40 & 50 & 47 \\			
				4  & 20 & 30 & 40 & 23 \\
				5  & 20 & 10 & 40 & 33 \\
				Mean & 30 & 28 & 42 & 33 \\
				\hline
			\end{tabular}
			
		\end{threeparttable}
	\end{center}
\end{table}

\begin{table}[t]
	\caption{Percent of Epochs with Two or More Wrong Detected/Missed Respiratory Cycles}
	\label{table_bad_epochs_2}
	\begin{center}
		\begin{threeparttable}[c]
			\begin{tabular}{c c c c c}
				\hline
				Subject & Supine, \% & Side, \% & Prone, \% & All Postures, \% \\
				\hline
				1  & 20 & 10 & 10 & 13 \\			
				2  & 0  & 0  & 0  & 0 \\
				3  & 0  & 0  & 20 & 7 \\			
				4  & 0  & 10 & 0  & 3 \\
				5  & 4  & 10 & 0  & 3 \\
				Mean & 30 & 6  & 6  & 5 \\
				\hline
			\end{tabular}
			
		\end{threeparttable}
	\end{center}
\end{table}

\begin{table}[t]
	\caption{Mean Absolute Difference Between Number of Detected Respiratory Cycles on CSI and Belt Signals during Epoch}
	\label{table_error}
	\begin{center}
		\begin{threeparttable}[c]
			\begin{tabular}{c c c c c c}
				\hline
				Subject & BMI & Supine  & Side & Prone & All Postures \\
				\hline
				1  & 31 & 0.60 & 0.70 & 0.80 & 0.70 \\			
				2  & 17 & 0.20 & 0.00 & 0.10 & 0.10 \\
				3  & 23 & 0.50 & 0.36 & 0.80 & 0.55 \\			
				4  & 25 & 0.20 & 0.40 & 0.40 & 0.33 \\
				5  & 34 & 0.20 & 0.20 & 0.40 & 0.27\\
				\hline
				\multicolumn{2}{c}{Mean} & 0.34 & 0.33 & 0.50 & 0.39\\
				\hline
			\end{tabular}
			\begin{tablenotes}
				\item	\begin{flushright} BMI --- Body Mass Index
				\end{flushright}
			\end{tablenotes}
		\end{threeparttable}
	\end{center}
\end{table}

\section{Discussion}

The aim of the research was to check whether it is possible to use Wi-FI CSI on a smartphone for breathing monitoring. To the best of our knowledge, this is the first study addressed to this question. Our results show that information about breathing is contained in smartphone's CSI and can be extracted for subjects with different BMI in different sleep postures. In spite of the fact that the results are quite preliminary, we can say that using CSI on smartphones for sleep and breathing monitoring is a promising direction for research.

Accurate and reliable detection of respiratory cycles and their parameters might be tricky due to high noise level in the signal. This is especially true for prone position, when CSI predominantly modulated by back movements with small amplitude. However, the dataset is too small to make reasonable conclusions about influence of BMI and postures on the signal quality. It is still a question whether the quality of the respiratory signal in the smartphone's CSI enough for sleep monitoring.

The further research should consider a larger and more diverse dataset with CSI collected during sleep together with PSG to evaluate eligibility of its use for sleep staging. Moreover, phase of CSI should be used in additional to amplitude. Wang et al.~\cite{Wang2017b} shows that the phase difference between two receiver's antennas is very useful for breathing monitoring. Thus, the smartphone with two or more antennas should be used to have advantage of using both amplitude and phase of CSI for breathing and sleep monitoring. 




\begin{thebibliography}{10}
\providecommand{\url}[1]{#1}
\csname url@samestyle\endcsname
\providecommand{\newblock}{\relax}
\providecommand{\bibinfo}[2]{#2}
\providecommand{\BIBentrySTDinterwordspacing}{\spaceskip=0pt\relax}
\providecommand{\BIBentryALTinterwordstretchfactor}{4}
\providecommand{\BIBentryALTinterwordspacing}{\spaceskip=\fontdimen2\font plus
\BIBentryALTinterwordstretchfactor\fontdimen3\font minus
  \fontdimen4\font\relax}
\providecommand{\BIBforeignlanguage}[2]{{%
\expandafter\ifx\csname l@#1\endcsname\relax
\typeout{** WARNING: IEEEtran.bst: No hyphenation pattern has been}%
\typeout{** loaded for the language `#1'. Using the pattern for}%
\typeout{** the default language instead.}%
\else
\language=\csname l@#1\endcsname
\fi
#2}}
\providecommand{\BIBdecl}{\relax}
\BIBdecl

\bibitem{Ong2016a}
A.~A. Ong and M.~B. Gillespie, ``{Overview of smartphone applications for sleep
  analysis},'' \emph{World Journal of Otorhinolaryngology-Head and Neck
  Surgery}, vol.~2, no.~1, pp. 45--49, 2016.

\bibitem{Kerkhof2017}
G.~A. Kerkhof, ``{Epidemiology of sleep and sleep disorders in The
  Netherlands},'' \emph{Sleep Medicine}, vol.~30, pp. 229--239, 2017.

\bibitem{Simonelli2018}
G.~Simonelli, N.~S. Marshall, A.~Grillakis \emph{et~al.}, ``{Sleep health
  epidemiology in low and middle-income countries: a systematic review and
  meta-analysis of the prevalence of poor sleep quality and sleep duration},''
  \emph{Sleep Health}, vol.~4, no.~3, pp. 239--250, 2018.

\bibitem{Senaratna2017}
C.~V. Senaratna, J.~L. Perret, C.~J. Lodge \emph{et~al.}, ``{Prevalence of
  obstructive sleep apnea in the general population: A systematic review},''
  \emph{Sleep Medicine Reviews}, vol.~34, no. October, pp. 70--81, 2017.

\bibitem{Ohayon2002a}
M.~M. Ohayon, ``{Epidemiology of insomnia: What we know and what we still need
  to learn},'' \emph{Sleep Medicine Reviews}, vol.~6, no.~2, pp. 97--111, 2002.

\bibitem{Redmond2006}
S.~J. Redmond and C.~Heneghan, ``{Cardiorespiratory-based sleep staging in
  subjects with obstructive sleep apnea},'' \emph{IEEE Transactions on
  Biomedical Engineering}, vol.~53, no.~3, pp. 485--496, mar 2006.

\bibitem{Long2014}
X.~Long, J.~Foussier, P.~Fonseca \emph{et~al.}, ``{Analyzing respiratory effort
  amplitude for automated sleep stage classification},'' \emph{Biomedical
  Signal Processing and Control}, vol.~14, no.~1, pp. 197--205, nov 2014.

\bibitem{Tataraidze2016}
A.~Tataraidze, L.~Korostovtseva, L.~Anishchenko \emph{et~al.},
  ``\BIBforeignlanguage{english}{{Sleep Architecture Measurement Based on
  Cardiorespiratory Parameters}},'' in \emph{\BIBforeignlanguage{english}{38th
  Annual International Conference of the IEEE Engineering in Medicine and
  Biology Society}}, vol.~53, Orlando, USA, 2016, pp. 3478--3481.

\bibitem{Tataraidze2017}
A.~Tataraidze, L.~Anishchenko, L.~Korostovtseva \emph{et~al.}, ``{Estimation of
  a priori probabilities of sleep stages: A cycle-based approach},''
  \emph{Proceedings of the Annual International Conference of the IEEE
  Engineering in Medicine and Biology Society, EMBS}, pp. 3745--3748, 2017.

\bibitem{Beattie2017}
Z.~Beattie, Y.~Oyang, A.~Statan \emph{et~al.}, ``{Estimation of sleep stages in
  a healthy adult population from optical plethysmography and accelerometer
  signals},'' \emph{Physiological Measurement}, vol.~38, no.~11, pp.
  1968--1979, 2017.

\bibitem{Liu2015}
X.~Liu, J.~Cao, S.~Tang \emph{et~al.}, ``{Wi-sleep: Contactless sleep
  monitoring via WiFi signals},'' \emph{Proceedings - Real-Time Systems
  Symposium}, vol. 2015-Janua, no. January, pp. 346--355, 2015.

\bibitem{Li2017}
F.~Li, C.~Xu, Y.~Liu \emph{et~al.}, ``{Mo-sleep: Unobtrusive sleep and movement
  monitoring via Wi-Fi signal},'' \emph{2016 IEEE 35th International
  Performance Computing and Communications Conference, IPCCC 2016}, 2017.

\bibitem{Gu2018a}
Y.~Gu, J.~Zhan, Z.~Liu \emph{et~al.}, ``{Sleepy: Adaptive sleep monitoring from
  afar with commodity WiFi infrastructures},'' \emph{IEEE Wireless
  Communications and Networking Conference, WCNC}, vol. 2018-April, pp. 1--5,
  2018.

\bibitem{roman6}
``{Cisco Aironet 3.5-dBi Articulated Dipole Antenna (AIR-ANT5135DW-R)},'' 2006.

\bibitem{roman7}
C.~Rowell and E.~Y. Lam, ``{Mobile-Phone Antenna Design},'' \emph{IEEE Antennas
  and Propagation Magazine}, vol.~54, no.~4, pp. 14--34, aug 2012.

\bibitem{roman10}
``{Dual-band Reach XtendTM (FR05-S1-NO-1-003) – 2.4-2.5 GHz and 4.9-5.875
  GHz},'' 2015.

\bibitem{roman12}
``{Compact Dual-band Reach Xtend TM ( FR05-S1-NO-1-004 ) –},'' 2015.

\bibitem{Davies2011}
L.~Davies and U.~Gather, ``{The Identification of Multiple Outliers},''
  \emph{Journal of the American Statistical Association}, vol.~88, no. 423, p.
  782, sep 1993.

\bibitem{Pearson1901}
K.~Pearson, ``{LIII. On lines and planes of closest fit to systems of points in
  space},'' \emph{The London, Edinburgh, and Dublin Philosophical Magazine and
  Journal of Science}, vol.~2, no.~11, pp. 559--572, nov 1901.

\bibitem{Wang2017b}
X.~Wang, C.~Yang, and S.~Mao, ``{PhaseBeat: Exploiting CSI Phase Data for Vital
  Sign Monitoring with Commodity WiFi Devices},'' \emph{Proceedings -
  International Conference on Distributed Computing Systems}, pp. 1230--1239,
  2017.

\end{thebibliography}

\end{document}